\def\e{{\rm e}}
\def\del{\partial}
\def\half{{1\over2}}

\def\vev#1{\langle #1 \rangle}

\def\del{\partial}
\def\dslash{\del\kern-0.55em\raise 0.14ex\hbox{/}}

\def\rough#1{\raise.3ex\hbox{$#1$\kern-.75em\lower1ex\hbox{$\sim$}}}

\newcommand{\hmu}{\hat\mu}

\newcommand{\bea}{\begin{eqnarray}}
\newcommand{\eea}{\end{eqnarray}}

\def\e{{\rm e}}
\def\del{\partial}
\def\half{{1\over2}}

\def\vev#1{\langle #1 \rangle}
\def\del{\partial}
\def\half{{1\over2}}

\def\vev#1{\langle #1 \rangle}

\def\del{\partial}
\def\dslash{\del\kern-0.55em\raise 0.14ex\hbox{/}}

\def\rough#1{\raise.3ex\hbox{$#1$\kern-.75em\lower1ex\hbox{$\sim$}}}

\newcommand{\tpsi}{{\tilde\psi}}
\documentclass[12pt]{article}
\usepackage[xdvi]{graphicx}
\makeatletter 

\@addtoreset{equation}{section}
\makeatother 
\begin{document}
\baselineskip=18pt
\begin{titlepage}
\begin{flushright}
ROMA-1431/06 \\
TU-772
\end{flushright}
\vspace*{2cm}
\begin{center}{\Large\bf Bulk Mass Effects 
in Gauge-Higgs Unification 
\\
[3mm]
at Finite Temperature}
\end{center}
\vspace{1cm}
\begin{center}
{\bf Nobuhito Maru}$^{(a)}$
\footnote{E-mail: Nobuhito.Maru@roma1.infn.it},
{\bf Kazunori Takenaga}$^{(b)}$
\footnote{E-mail: takenaga@tuhep.phys.tohoku-u.ac.jp}
\end{center}
\vspace{0.2cm}
\begin{center}
${}^{(a)}$ {\it Dipartimento di Fisica, 
Universit\`a di Roma "La Sapienza" \\and INFN, Sezione di Roma, 
P.le Aldo Moro, I-00185 Roma, Italy}
\\[0.2cm]
${}^{(b)}$ {\it Department of Physics, Tohoku University, 
Sendai 980-8578, Japan}
\end{center}
\vspace{1cm}
\begin{abstract}
We study the bulk mass effects on the electroweak phase transition 
at finite temperature in a five dimensional 
$SU(3)$ gauge-Higgs unification model on an orbifold. 
We investigate whether the Higgs mass 
satisfying the experimental lower bound can be compatible with 
the strong first order phase transition necessary 
for a successful electroweak baryogenesis. 
Our numerical results show that the above statement can be realized 
by matter with bulk mass yielding a viable Higgs mass. We also
find an interesting case where the heavier Higgs gives the
stronger first order phase transition.
\end{abstract}
\end{titlepage}
\newpage
\section{Introduction}
Solving the gauge hierarchy problem is one of the guidelines 
to consider the physics beyond the Standard Model. 
Low energy supersymmetry is well motivated theory to solve 
the gauge hierarchy problem. 
Furthermore, it predicts the gauge coupling unification and 
the natural candidate for the dark matter. 
However, the superparticle has not yet been discovered so far. 
Therefore, it is worth while to explore other possibilities 
beyond the Standard Model. 
Recent developments of the physics of extra dimensions offer 
many new insights into the four dimensional physics 
including the gauge hierarchy problem.

The gauge-Higgs unification \cite{gaugehiggs1}-\cite{ABQ} 
is one of the attractive scenarios of this kind. 
In this scenario, the Higgs scalar is originated from the extra component of 
the gauge field in higher dimensions and the gauge symmetry breaking occurs 
by the vacuum expectation value (VEV) through the dynamics of 
the Wilson line phases called as Hosotani mechanism \cite{hosotani}. 
Now, the gauge-Higgs unification has been widely studied 
from the various viewpoints \cite{gaugehiggs3}-\cite{LMH}. 

Remarkable thing is that the Higgs mass at the classical level 
is forbidden by the higher dimensional gauge symmetry. 
Furthermore, the Higgs mass is generated by quantum corrections 
and insensitive to the cutoff scale. 
This is an essential feature of solving the gauge hierarchy problem. 
In fact, explicit calculations showing the finiteness of the Higgs mass 
are known for the five dimensional(5D) QED compactified on $S^1$ 
at one-loop level \cite{HIL} and two-loop level\cite{MY}, 
5D non-Abelian gauge theory on an orbifold $S^1/Z_2$ at one-loop level 
\cite{GIQ}, and 6D $U(3) \times U(3)$ gauge theory on toroidal 
compactification at one-loop 
level \cite{ABQ} \cite{hosont}, and 6D massive scalar QED on $S^2$ 
at one-loop level \cite{LMH}. 
Even in the Gravity-Gauge-Higgs unification scenario, 
the Higgs mass is shown to be finite at one-loop level 
in 5D gravity theory on $S^1$coupled to the bulk scalar field \cite{HLM}.

However, it is highly nontrivial task to obtain the phenomenologically 
viable Higgs mass in such a scenario because the Higgs quartic coupling 
is also generated at one-loop level, which implies that 
the Higgs mass is generically too small. 
Recent works for this problem are found in 
\cite{CCP} in terms of large representation fermion contributions and 
\cite{PSW} in terms of the explicit 5D Lorentz symmetry breaking effects. 
In our previous paper \cite{MT2}, 
we have proposed another solution of this problem. 
We have investigated the bulk mass effects on 
the electroweak symmetry breaking and the size of Higgs mass. 
The bulk mass of matter fields is a necessary ingredient 
in gauge-Higgs unification to discuss Yukawa hierarchy \cite{yukawa} 
and the pattern of the gauge symmetry breaking \cite{HTY}.
We have found some simple matter content satisfying 
the lower bound of Higgs mass where the fields with an antiperiodic 
boundary condition play an important role to make the Higgs VEV small.

On the other hand, one of the interesting applications of 
gauge-Higgs unification is the electroweak baryogenesis 
at finite temperature \cite{PS, MT}. 
The dynamics of Wilson line phases is essentially the Coleman-Weinberg 
mechanism \cite{CW}, 
so that the phase transition at finite temperature is expected 
to be the first order \cite{FKT}. 
As will be discussed later, 
the first order phase transition is controlled 
by the cubic term of the effective potential, 
which originates from the zero mode of bosonic fields. 
Comparing to the four dimensional case, 
the contributions from the higher dimensional gauge field 
is crucial to strengthen the first order phase transition.

In \cite{MT}, 
we have studied the finite temperature behavior 
of the 5D $SU(3)$ gauge-Higgs unification models 
with massless bulk matter in adjoint and the fundamental 
representations. In fact, we have confirmed that the 
phase transition is indeed 
the first order and found some examples of matter content 
satisfying the Higgs mass 
lower bound and the strong first order transition 
necessary for the baryogenesis. 
As stated above, 
since the bulk mass is indispensable in gauge-Higgs unification, 
it is worth while to study the bulk mass effects on 
the electroweak phase transition at finite temperature.

From these motivations, in this paper, 
we investigate the gauge-Higgs unification with massive bulk matter 
at finite temperature 
and focus on the possibility that the Higgs mass satisfying 
the experimental lower bound and 
the strong first order phase transition required 
for successful electroweak baryogenesis are compatible. 
Due to the general properties that 
the large dimensional representation fermions weaken 
the first phase transition \cite{DJ}, 
we consider the bulk matter 
with only the fundamental and 
the adjoint representations as in \cite{MT}. We 
find that the matter content considered in \cite{MT2} have 
a strong first order phase transition enough for 
a successful electroweak baryogenesis. And we also find 
an interesting case where the heavier Higgs mass has the
stronger first order phase transition.

This paper is organized as follows. 
In the next section, we derive the one-loop effective potential 
at finite temperature for the Higgs field. 
The results obtained in \cite{MT2} for the analysis 
of the Higgs mass is reviewed. Section 3 
is the main part of this paper, where the bulk mass effects 
on the electroweak phase transition at finite temperature is investigated 
using the models studied in \cite{MT2}. 
Section 4 is devoted to the conclusions of this paper. 

\section{Effective potential of the model}
Let us study the effect of the finite temperature 
on the gauge-Higgs unification. 
Namely, we are interested in the order of the phase transition 
caused by the dynamics of the Wilson line (Hosotani mechanism). 
We take the $D$ dimensional space-time to be
\begin{equation}
M^D =(S^1 \times R^{D-2}) \times S^1/Z_2,
\end{equation}
where one of the spatial coordinate is compactified on an orbifold, 
$S^1/Z_2$ and the Euclidean time direction corresponds to a circle $S^1$.
Accordingly, the gauge potential is decomposed as 
\begin{equation}
A_{\hmu} \equiv A_{\hmu}^a T^a = (A_0, A_i, A_y) 
\qquad  (i=1,\cdots,D-2),
\end{equation}
where $T^a$ is the generators of the gauge group we are considering with 
the normalization being $\mbox{tr}(T^aT^b)=\half \delta_{ab}$.

We consider the $SU(3)$ gauge group as the simplest 
example of the gauge-Higgs unification. 
One of the spatial coordinate is an $S^1/Z_2$, so that there are two fixed
points at $y=0,\pi R$, where $R$ is the radius of $S^1$. 
Considering the gauge theory on such a space-time, one needs to
specify the boundary conditions of fields for the $S^1$ direction 
and the fixed point. Here we define them as
\begin{eqnarray}
A_{\hmu}(x, y + 2\pi R) &=&
U A_{\hmu}(x, y ) \, U^\dagger ,\\
\pmatrix{A_\mu \cr A_y \cr} (x, y_i - y) &=&
P_i \pmatrix{A_\mu \cr - A_y \cr} (x, y_i + y) \, 
P_i^\dagger,~~ (i = 0, 1)
\end{eqnarray}
where $U^\dagger = U^{-1}, P_i^\dagger = P_i= P_i^{-1}$ and 
$y_0=0, y_1=\pi R$. The minus sign for $A_y$ is needed to preserve 
the gauge invariance under these transformations. 
Since a transformation $\pi R+y \rightarrow \pi R -y$ must be 
the same as a transformation 
$\pi R +y \rightarrow -(\pi R + y)\rightarrow \pi R -y$, 
we obtain 
\begin{equation}
U = P_1 P_0. 
\end{equation}  
Hereafter, we consider $P_i$ to be fundamental quantities.
\par
The gauge symmetry at low energies is determined by counting the zero
modes associated with $A_{\mu}$ for each $P_{0,1}$ at the tree level. 
We are interested in how the electroweak gauge symmetry, 
$SU(2)\times U(1)$ is broken at the quantum level. 
Therefore, we break the original $SU(3)$ gauge symmetry 
by the orbifolding down to $SU(2)\times U(1)$ 
by taking $P_0=P_1=\mbox{diag}(-1,-1,1)$. 
Then, the zero mode are found to be 
\begin{eqnarray}
A_{\mu}^{(0)} &=& \half \pmatrix{
A_{\mu}^3 +{A_{\mu}^8\over\sqrt{3}} & A_{\mu}^1-iA_{\mu}^2  & 0 \cr 
A_{\mu}^1+iA_{\mu}^2 & -A_{\mu}^3+{A_{\mu}^8\over\sqrt{3}}  & 0 \cr
0 & 0  & -{2\over\sqrt{3}}A_{\mu}^8 },\\
A_y^{(0)}&= &\half \pmatrix{
0& 0 & A_y^4-iA_y^5 \cr 
0& 0  & A_y^6-iA_y^7 \cr
A_y^4+iA_y^5 & A_y^6+iA_y^7 & 0 }.
\end{eqnarray}
One can observe that the zero mode of $A_y$
\begin{equation}
\Phi \equiv \pmatrix{A_y^4 -i A_y^5 \cr A_y^6 -i A_y^7 \cr} 
\end{equation}
transforms as a doublet under the $SU(2)$ gauge symmetry, 
so that we can regard $\Phi$ as the Higgs doublet in the Standard Model.  
\par
By using the $SU(2) \times U(1)$ degrees of freedom, 
the VEV of $A_y$ $(\vev{A_y})$ can be parameterized as
\begin{equation} 
\vev{A_y} = {a\over{gR}}{\lambda^6\over 2},
\end{equation}
where $\lambda^6$ is the sixth Gell-Mann matrix and $a$ is a 
dimensionless real parameter. 
$g$ is the five dimensional gauge coupling whose mass dimension is $-1/2$. 
The parameter $a$ 
is closely related to the Wilson line phases as follows, 
\begin{eqnarray}
W &=& {\cal P} \mbox{exp} \left(ig \oint_{S^1}dy A_y \right)
= \pmatrix{
1 & 0 & 0 \cr 
0 & \cos(\pi a)  & i\sin(\pi a) \cr
0 & i\sin(\pi a) & \cos(\pi a) }\qquad (\mbox{mod}~2)
\nonumber\\
&=& \left\{
\begin{array}{llll}
\mbox{diag}(1,1,1) & \mbox{for} & a=0,  & SU(2)\times U(1),\\
\mbox{diag}(1,-1,-1)& \mbox{for} & a=1, & U(1)^{\prime}\times U(1).
\end{array}\right.
\end{eqnarray}
We observe that the gauge symmetry breaking patterns are classified 
by the values of $a$ and in order to obtain 
the desirable electroweak symmetry breaking 
$SU(2) \times U(1) \rightarrow U(1)$, 
one needs the fractional values of $a$. 
As we will see later, the values of $a$ is determined dynamically 
through the dynamics of the Wilson line phases.

As mentioned in our previous paper \cite{MT2}, 
we note that the bulk mass term of fermions in five dimensions 
is odd under the parity transformation $y \to -y(\pi R +y \to \pi R -y)$. 
We introduce a pair of fields $\psi_\pm$ whose parity is given by 
$\psi_\pm (-y) = \pm \psi_\pm(y)$ to obtain the mass term like 
$M \bar{\psi}_+ \psi_-$ where $M$ is a constant. Let us 
suppose that $\psi^{(\pm)}$ and ${\tilde\psi}^{(\pm)}$ belong to
the fundamental representation under the $SU(3)$ gauge group and satisfy
the following boundary conditions \footnote{Notations used in this 
section is the same as those in \cite{HTY}.},
\begin{eqnarray}
\mbox{type~I}&;&\left\{
\begin{array}{l}
\psi^{(\pm)}(-y)=\pm P_0~i\Gamma^y~\psi^{(\pm)}(y)\\[0.3cm]
\psi^{(\pm)}(\pi R-y)=\pm P_1~i\Gamma^y~\psi^{(\pm)}(\pi R+ y),
\end{array}\right.\label{shiki4}
\\
\mbox{type~II}&;&\left\{
\begin{array}{l}
\tpsi^{(\pm)}(-y)=\pm P_0~i\Gamma^y~\tpsi^{(\pm)}(y)\\[0.3cm]
{\tpsi}^{(\pm)}(\pi R-y)=\mp P_1~i\Gamma^y~{\tpsi}^{(\pm)}(\pi R+y).
\end{array}\right.
\label{shiki5}
\end{eqnarray}
Equipped with these fields, a pair $(\psi^{(+)}, \psi^{(-)}) ((\tpsi^{(+)},
\tpsi^{(-)})$ can have the parity
even and gauge invariant bulk mass term,
$M{\bar\psi}^{(\pm)}\psi^{(\mp)} (M{\bar\tpsi}^{(\pm)}\tpsi^{(\mp)})$.
It is easy to see that the 
field $\psi^{\pm}$ satisfies the periodic boundary 
condition, $\psi^{\pm}(y+L)=\psi^{\pm}(y)$,
while ${\tilde\psi}^{\pm}$ does the antiperiodic boundary
condition, ${\tilde\psi}^{\pm}(y+L)=-{\tilde\psi}^{\pm}(y)$.

Let us first calculate the bosonic effective 
potential at finite temperature 
for the real parameter $a$. The finite temperature
part of the effective potential is taken into account
by compactifying the Euclidean time direction on $S^1$ with
period being $\beta=T^{-1}$, where $T$ is the temperature. And
accordingly, the momentum for the time direction is discretized as 
\begin{equation}
p_{\tau} = 2\pi l \times T,\qquad l\in {\bf Z}.  
\end{equation}
Hence, the effective potential we study is
\begin{equation}
V_B^T = N_{{\rm deg}}^{\rm B} {T \over {2L}} 
\sum_{l,n=-\infty}^{\infty} \int 
{{d^{D-2}p_E}\over{(2\pi)^{D-2}}}{\rm ln}
\Biggl[p_{E(D-2)}^2 + (2\pi T )^2 l^2 + 
\left( {{n+Qa-{\delta\over 2}} \over R}\right)^2 + M^2 \Biggr],
\label{boson2}
\end{equation}
where $N_{{\rm deg}}^{\rm B}$ denotes the on-shell 
degrees of freedom of the bosonic fields we 
concern. The charge $Q$ is $ \half $ for matter belonging to the 
fundamental representation 
under the gauge group, and for other representation the
expression is given by the combination of Eq. (\ref{boson2}) with 
the suitable values of $Q$. And $\delta$ 
takes $0$ or $1$, depending on the periodicity for the
$S^1$ direction \cite{MT2}. In 
Eq.(\ref{boson2}), the 4D momentum is taken to be Euclidean. 
$M$ stands for the bulk mass of the matter fields 
which was not considered in the previous paper \cite{MT}.

On the other hand, as for fermions, according to the quantum statistics, 
since the fermion must take the anti-periodic boundary condition 
for the Euclidean time direction, 
contrary to the boson taking the periodic one,
we have 
\begin{equation}
p_{\tau} = 2\pi \left( l+\half \right) \times T.
\end{equation}
Hence, for fermions we study 
\begin{eqnarray}
V_F &=& -2^{\left[\frac{D}{2}\right]}(2N^{pair}){T \over {2L}} 
\sum_{l,n=-\infty}^{\infty}
\int {{d^{D-2}p_E}\over{(2\pi)^{D-2}}}\nonumber\\
&\times &{\rm ln}
\Biggl[p_{E(D-2)}^2 + (2\pi T)^2(l+\half)^2 
+ \left( {{n+Qa-{\delta\over 2}} \over R}\right)^2 + M^2 \Biggr],
\label{fermion}
\end{eqnarray}
where it is understood that $\delta$ takes $0$ or $1$.
\par
We can separate the zero temperature $(l=0)$ and the finite
temperature parts 
as $V_{B(F)}=V_{B(F)}^{T=0}+V_{B(F)}^{T\neq 0}$. We regularize 
the zero temperature part of the
effective potential by 
subtracting the $n=0$ mode. On the other hand, the remaining 
finite temperature part of the effective potential has no 
divergence thanks to the Boltzmann suppression factor associated with 
the finite temperature. Following the usual prescription \cite{DJ}, 
we obtain the effective potential for bosons 
\bea
\label{0Tboson}
V_B^{T=0} &=& - N_{{\rm deg}}^{\rm B} \frac{2}{(2\pi)^{\frac{D}{2}}}
\sum_{n=1}^{\infty} \left( \frac{M}{nL} \right)^{\frac{D}{2}}
K_{\frac{D}{2}}(nLM)\cos \left[ 2\pi n \left(Qa - {\delta\over 2} \right)
\right], \\
V_B^{T \ne 0} &=& -N_{{\rm deg}}^{\rm B} 
\frac{4}{(2\pi)^{\frac{D}{2}}} 
\sum_{l=1}^{\infty} \sum_{n=1}^{\infty} 
\left( \frac{M^2 T^2}{(nLT)^2+l^2} \right)^{\frac{D}{4}}
K_{\frac{D}{2}}\left( \frac{M}{T}\sqrt{(nLT)^2+l^2} \right) 
\nonumber \\
&&\times \cos \left[ 2\pi n \left(Qa - {\delta\over 2} \right) \right],
\label{bosonpot}
\eea
where $K_n(x)$ is the modified Bessel function. Similarly, the 
fermion contribution to the effective potential 
at finite temperature is calculated as
\begin{eqnarray}
V_F &=& V_F^{T=0}+V_F^{T\neq 0}\nonumber\\
&=&
2^{\left[\frac{D}{2} \right]}(2N^{pair}) \frac{2}{(2\pi)^{\frac{D}{2}}}
\left[ 
\sum_{n=1}^{\infty} \left( \frac{M}{nL} \right)^{\frac{D}{2}}
K_{\frac{D}{2}}(LMn) \right. \nonumber \\
&& \left. +2\sum_{l=1}^{\infty} \sum_{n=1}^{\infty} (-1)^l
\left( \frac{M^2 T^2}{(nLT)^2+l^2} \right)^{\frac{D}{4}}
K_{\frac{D}{2}}\left( \frac{M}{T}\sqrt{(nLT)^2+l^2} \right) \right]
\cos \left[ 2\pi n \left(Qa - {\delta\over 2} \right) \right]. 
\nonumber \\
\label{fermionpot}
\end{eqnarray}
\subsection{Review of Higgs mass results}
In this subsection, we focus on the model of an $SU(3)$ gauge theory 
in five dimensions introduced in the previous section. 
For completeness, we briefly review the numerical results of Higgs mass 
studied in the previous paper \cite{MT2}. 
Then, the phase transition at finite temperature will be investigated 
in the next section using the same models reviewed here.

Following the definition in \cite{MT2}, 
we introduce the flavor numbers specifying the matter content as
\bea
(N_{adj}^I,N_{fd}^I, N_{adj}^{(+)s}, N_{fd}^{(+)s};
N_{adj}^{II}, N_{fd}^{II}, N_{adj}^{(-)s}, N_{fd}^{(-)s}) 
\label{shiki13}
\eea
where the matter representation under $SU(3)$ gauge group 
is the adjoint or the fundamental representations. 
The index $s$ means the scalar field. $I,II$ denote 
the type of the boundary conditions for fermions defined 
in (\ref{shiki4}) and (\ref{shiki5}). 
$(+),(-)$ denotes the $\eta$-parity of the scalar field 
which is essentially same as the periodic, antiperiodic 
field, respectively.

Let us consider the following two cases 
and list up the corresponding results in 
Table \ref{A} and \ref{B} which were obtained 
in \cite{MT2}\footnote{We simply dropped the result 
of the case (C) in \cite{MT2} and 
its finite temperature behavior because the 
qualitative behaviors are almost the same 
as that of the case (A).}. 
\begin{eqnarray}
(\mbox{A})~~(N_{adj}^I,N_{fd}^I, N_{adj}^{(+)s}, N_{fd}^{(+)s}; N_{adj}^{II}, 
N_{fd}^{II}, N_{adj}^{(-)s}, N_{fd}^{(-)s})&=&(1,1,0,0;1,1,1,0), 
\label{matterA}\\
(\mbox{B})~~(N_{adj}^I,N_{fd}^I, N_{adj}^{(+)s}, N_{fd}^{(+)s}; N_{adj}^{II}, 
N_{fd}^{II}, N_{adj}^{(-)s}, N_{fd}^{(-)s})&=&(1,1,0,2;1,1,2,0).
\label{matterB}
\end{eqnarray}
\begin{table}[t]
$$
\begin{array}{|c|cccc|ccccccc|}
\hline
 & z_{adj}^{(+)} & z_{fd}^{(+)} & z_{adj}^{(+)s} & z_{fd}^{(+)s} &  
   z_{adj}^{(-)} & z_{fd}^{(-)}   & z_{adj}^{(-)s} & z_{fd}^{(-)s}& 
   {1\over{g_4R}} & a_0 & m_H/g_4^2 
\\ \hline\hline
(1)& 0 & 0 & \mbox{-} & \mbox{-} & 0 & 0 & 0 & \mbox{-} & 6.3
& 0.039 & 134.0 
\\ \hline
(2)& 0.1 & 0.2 & \mbox{-} & \mbox{-} & 0 & 0 & 0 & \mbox{-} & 12.4
& 0.020 & 139.4
\\ \hline
(3)& 0 & 0 & \mbox{-} & \mbox{-} & 0 & 0 & 0.5 & \mbox{-} & 7.7
& 0.032 & 141.0
\\ \hline
(4)& 0.1 & 0.2 & \mbox{-} & \mbox{-} & 0 & 0 & 0.2 & \mbox{-} & 14.0
& 0.018 & 140.4
\\ \hline
(5)& 0.1 & 0.2 & \mbox{-} & \mbox{-} & 0.1 & 0 & 0 & \mbox{-} & 10.4
& 0.024 & 137.2
\\ \hline
(6)& 0.1 & 0.2 & \mbox{-} & \mbox{-} & 0 & 0.2 & 0 & \mbox{-} & 11.0
& 0.022 & 137.9
\\ \hline
(7)& 0.2 & 0.2 & \mbox{-} & \mbox{-} & 0.3 & 0.3 & 0 & \mbox{-} & 24.0
& 0.010 & 109.1
\\ \hline
\end{array}
$$
\caption{Higgs mass results of the matter content (A). 
The dimensionless parameter $z$ is defined as $z \equiv ML$.}
\label{A}
\end{table}
\begin{table}[t]
$$
\begin{array}{|c|cccc|ccccccc|}
\hline
 & z_{adj}^{(+)} & z_{fd}^{(+)} & z_{adj}^{(+)s} & z_{fd}^{(+)s} &  
   z_{adj}^{(-)} & z_{fd}^{(-)}   & z_{adj}^{(-)s} & z_{fd}^{(-)s}& 
   {1\over{g_4R}} & a_0 & m_H/g_4^2 
\\ \hline\hline
(1)& 0 & 0 & \mbox{-} & 0 & 0 & 0 & 0 & \mbox{-} & 4.0
& 0.062 & 117.4 
\\ \hline
(2)& 0.2 & 0.2 & \mbox{-} & 0.2 & 0 & 0 & 0 & \mbox{-} &10.4
& 0.024 & 118.6
\\ \hline
(3)& 0.2 & 0.1 & \mbox{-} & 0.2 & 0 & 0 & 0 & \mbox{-} & 7.7
& 0.032 & 120.0
\\ \hline
(4)& 0.2 & 0.25 & \mbox{-} & 0.2 & 0 & 0 & 0 & \mbox{-} & 15.5
& 0.016 & 115.0
\\ \hline
(5)& 0.2 & 0.3 & \mbox{-} & 0.5 & 0 & 0 & 0 & \mbox{-} & 7.5
& 0.033 & 118.2
\\ \hline
\end{array}
$$
\caption{Higgs mass results of the matter content (B).}
\label{B}
\end{table}
We discussed in \cite{MT2} that the viable Higgs mass
is obtained by an interplay of the bulk mass effects between 
the periodic and the antiperiodic fields based on the
transparent and useful expression for the effective potential. 
We also showed, using the expression, that the size of 
the Higgs mass is mainly controlled 
by the periodic field contributions, but the antiperiodic 
field contributions also play an important role 
to obtain the small Higgs VEV necessary for the heavy Higgs mass. 
In the two Tables, the Higgs mass is measured in GeV and the radius
of $S^1$ is in TeV.

It would be interesting to study whether the above results are 
compatible with the electroweak baryogenesis as one of the applications 
of the gauge-Higgs unification at finite temperature. 
This is the main issue of this paper and will be investigated 
in the next section.

\section{Phase Transition at Finite Temperature}
One of the interesting applications of gauge-Higgs 
unification at finite temperature is the 
electroweak baryogenesis \cite{PS, MT}. It is known that 
the strong first order phase transition is required 
for the electroweak baryogenesis to work well. 
In the Standard Model, although there is a parameter 
space to realize the strong first order phase 
transition, the upper bound of Higgs mass becomes below 
the lower bound from LEP experiment. Therefore, the 
electroweak baryogenesis does not work in the standard model. 
In our previous work \cite{MT}, we have investigated 
the electroweak phase transition at finite temperature in the context of 
gauge-Higgs unification. We have found that some models have actually 
give rise to the strong first order phase transition compatible with 
the lower bound of Higgs mass. In this analysis, we have only 
considered massless bulk fields.
In gauge-Higgs unification, however, 
the massive bulk fields are necessary ingredients 
for generating Yukawa hierarchy \cite{yukawa} and give 
nontrivial effects on the pattern of the 
gauge symmetry breaking \cite{HTY}. Therefore, it is 
very important to investigate the effects of the 
bulk mass on the electroweak phase transition.

The effective potential at finite temperature we study 
is obtained from (\ref{bosonpot}) and (\ref{fermionpot}) 
\begin{eqnarray}
{\bar V}_{{\rm eff}}(a) = {V_{{\rm eff}} \over {4/(2 \pi)^{5/2}}} &=&
-3\biggl(f^T(2a, 0, 0) +2f^T(a,0, 0)\biggr)\nonumber\\
&+&4(2N_{adj}^I)\biggl(f^T(2a, z_{adj}^{(+)}, 0) + 
2f^T(a, z_{adj}^{(+)}, 0)\biggr) 
\nonumber\\
&+&4(2N_{adj}^{II})\biggl(f^T(2a, z_{adj}^{(-)}, 1) 
+ 2f^T(a, z_{adj}^{(-)}, 1)\biggr) 
\nonumber\\
&+&4(2N_{fd}^I) f^T(a, z_{fd}^{(+)}, 0) 
+4(2N_{fd}^{II})f^T(a, z_{fd}^{(-)}, 1) \nonumber\\
&-&dN_{adj}^{(+)s}\biggl(f^T(2a, z_{adj}^{(+)s}, 0) 
+ 2f^T(a, z_{adj}^{(+)s}, 0)\biggr) 
\nonumber\\
&-&dN_{adj}^{(-)s}\biggl(f^T(2a, z_{adj}^{(-)s}, 1) 
+ 2f^T(a, z_{adj}^{(-)s}, 1)\biggr) 
\nonumber\\
&-&(2N_{fd}^{(+)s}) f^T(a, z_{fd}^{(+)s}, 0) 
-(2N_{fd}^{(-)s})f^T(a, z_{fd}^{(-)s}, 1)
\label{finiteTpot}
\end{eqnarray}
where 
\bea
f^T(Qa, z, \delta) &=& \sum_{l=1}^\infty \sum_{n=1}^\infty (-1)^{lF}
\left( \frac{M^2 T^2}{(nLT)^2+l^2} \right)^{5/4} \nonumber \\
&& \times K_{5/2}\left( \frac{z}{LT}\sqrt{(nLT)^2+l^2} \right) 
\cos \left[ 2\pi n \left( Qa - \frac{\delta}{2} \right) \right] 
\eea
where $z \equiv ML$. 
In order to obtain the first order phase transition, 
the cubic term with respect to the order parameter $a$ in 
the potential plays an essential role. 
It is instructive to see where the cubic term comes 
from \cite{PS}\cite{DJ}. For that purpose, we 
approximate the above potential at high temperature. 
It is useful to start from the following expression which is obtained 
by using the zeta function regularization and 
$A^{-s}=\frac{1}{\Gamma(s)}\int_0^\infty dt t^{s-1}\e^{-At}$, 
\bea
V_{{\rm eff}}^{T \ne 0} &=& (-1)^{F+1}N_{{\rm deg}}N_{{\rm flavor}}
\frac{T}{2L} \sum_{l,n}\frac{\pi^{\frac{D-2}{2}}}{(2 \pi)^{D-2}} 
\nonumber \\
&&\times \int_0^\infty dt t^{-D/2} {\rm exp} \left[-\left\{
(2\pi T)^2(l+\eta)^2 + \left( \frac{n + Qa - \frac{\delta}{2}}{R} \right)^2 
+ M^2 \right\}t
\right]
\label{DpotnonT}
\eea
where $F=0(1)$ for bosons (fermions) and $\eta=0(1/2)$ for 
bosons (fermions). $N_{{\rm flavor}}$ means 
$N_{adj(fd)}^{(\pm)s}$ for bosons, and $2N_{adj(fd)}^{(I, II)}$
for fermions. Taking the Poisson resummation with respect to $n$, 
we obtain
\bea
&&\sum_{n=-\infty}^{\infty}
\int_0^\infty dt~t^{-D/2} {\rm exp} 
\left[-\left\{(2\pi T)^2(l+\eta)^2 + M^2 \right\} t \right] 
\nonumber\\
&&\times
R \sqrt{\frac{\pi}{t}}
{\rm exp}\left[ -\frac{1}{4t} (nL)^2 
-2\pi i n  \left(Qa - \frac{\delta}{2} \right) \right] 
\nonumber \\
&=&\sum_{n=1}^{\infty}
2R\sqrt{\pi} \int_0^\infty dt~t^{-(D+1)/2} {\rm exp} \left[-\left\{
(2\pi T)^2(l+\eta)^2 + M^2 \right\} t -\frac{(nL)^2}{4t} \right]
\nonumber\\
&&\times
\cos \left[ 
2\pi n \left( Qa - \frac{\delta}{2}
\right) \right],
\label{Poisson}
\eea
where we have simply ignored the $n=0$ to remove the irrelevant
divergence. If we assume $T \gg M$, the most 
dominant part of the potential 
at high temperature is 
$l=0$ and $\eta=0$ mode. 
This implies that the dynamics of the phase transition is 
dominantly controlled by bosons. 
Then, the above integral is reduced to 
\bea
4R\sqrt{\pi} 
\left[
\frac{4M^2}{(nL)^2}
\right]^{\frac{D-1}{4}} K_{\frac{D-1}{2}}(nLM)
\cos \left[ 
2\pi n \left( Qa - \frac{\delta}{2} \right)
\right],
\eea
so that we obtain for high temperature that 
\bea
V_{{\rm B}}^{{\rm high~T}} \simeq (-1)N_{{\rm deg}}N_{{\rm flavor}} 
\frac{2TM^{\frac{D-1}{2}}}
{(2\pi)^{\frac{D-1}{2}}}\sum_{n=1}^\infty 
\frac{K_{\frac{D-1}{2}}(nLM)}{(nL)^{\frac{D-1}{2}}}
\cos\left[2\pi n \left( Qa - \frac{\delta}{2} \right)
\right].
\eea
In the five dimensional case under consideration, 
the following formula can be applied,
\bea
K_{n(={\rm integer})}(x) &=& \frac{1}{2}
\sum_{k=0}^{n-1} (-1)^k\frac{(n-k-1)!}
{k!(x/2)^{n-k}} \nonumber \\
&+&(-1)^{\eta+1}\sum_{k=0}^\infty \frac{(x/2)^{n+2k}}{k!(n+k)!}
\nonumber\\
&\times &
\left[ \ln \left(\frac{x}{2} \right) -\frac{1}{2} \psi(k+1) 
-\frac{1}{2} \psi(n+k+1) \right]
\eea
where 
\bea
\psi(x=1) &=& -\gamma_E~(\gamma_E:{\rm Euler~number}),~~
\psi(x=2) = 1-\gamma_E, \\
\psi(x=n \ge 3) &=& \sum_{r=1}^{n-1} \frac{1}{r} -\gamma_E. 
\eea
Then, we expand the modified Bessel function using this formula,
\bea
&&\sum_{n=1}^{\infty}\frac{K_2(nLM)}{(nL)^2}
\cos \left( 
2\pi n \left( Qa - \frac{\delta}{2} \right)
\right) 
\simeq
\frac{2}{M^2}\sum_{n=1}^{\infty}\frac{1}{(nL)^4}
\cos \left( 
2\pi n \left( Qa - \frac{\delta}{2} \right)
\right) \nonumber \\
&&+(-1)^{F+1}\frac{1}{M}\sum_{n=1}^{\infty}\frac{1}{(nL)^3}
\cos \left( 
2\pi n \left( Qa - \frac{\delta}{2} \right)
\right) + \cdots
\eea
which means that the first term with $\delta=0$ includes 
the cubic term in $a$ to realize the first order phase transition 
because of the formula
\bea
\sum^{\infty}_{n=1} \frac{\cos nx}{n^4} = \frac{1}{48}
\left[ 2\pi^2 (x-\pi)^2 -(x-\pi)^4 -\frac{7\pi^4}{15} \right]
~(0 \le x \le2 \pi).
\eea 
Let us note that, as expected, the bulk mass term does not 
affect the cubic term in the high temperature expansion.

On the other hand, there appears no cubic term from fermion fields. 
The fermion part of the effective potential can be obtained 
from (\ref{DpotnonT}) and (\ref{Poisson}), 
\begin{eqnarray}
V_F &=& (-1)^{1+1} N_{{\rm deg}}N_{{\rm flavor}} 
{T\over{2\pi R}}
{1\over 2}{{\pi^{{D-2}\over 2}}\over{(2\pi)^{D-2}}}
\int_0^{\infty}dt~t^{-D/2} 
\sum_{l,n}R\left({\pi\over t}\right)^{1/2}
\nonumber\\
&\times &{\rm exp}\left({-{{(\pi R n)^2}\over t}- 
(2\pi T)^2(l+\half)^2t -M^2t-2\pi i n (Qa-{\delta\over 2})}
\right)
\end{eqnarray}
One again needs the regularization for $n=0$ mode, but this is just
carried out by subtracting the $n=0$ mode formally. 
Then, we obtain 
\begin{eqnarray}
V_F &=& 
(-1)^{1+1}N_{{\rm deg}}N_{{\rm flavor}}
{{2T}\over{(2\pi)^{{D-1}\over 2}}}{1\over L^{D-1}}
\sum_{n=1}^{\infty}\sum_{l=-\infty}^{\infty}
\Biggl[(Ln)^2\biggl((2\pi T)^2(l+\half)^2+M^2\biggr)\Biggr]^{{D-1}\over 4}
\nonumber\\
&\times& K_{{D-1}\over 2}
\Biggl(\sqrt{(Ln)^2
\biggl((2\pi T)^2(l+\half)^2+M^2\biggr)}\Biggr)
\cos\left[ 2\pi n \left( Qa-{\delta\over 2} \right) \right]. 
\end{eqnarray}
There is no zero Matsubara mode (with respect to $l$) for the
fermion, so that the fermion always suppresses the effective
potential at high temperature. 
Here, we assume $LT \gg 1$ and use the formula given by
\begin{equation}
K_{\nu}(x)\simeq \sqrt{\pi\over {2x}}
\e^{-x}\Biggl[1+{{4\nu^2 -1}\over {8x}}+\cdots\Biggr].
\label{formula}
\end{equation}
We set $D=5$ and keep only the first term in the above formula,
and we sum up only the Matsubara mode $l=0, -1$ alone. 
Then we arrive at 
\begin{equation}
V_F\simeq (-1)^{1+1}N_{{\rm deg}}N_{{\rm flavor}}
 {T\over {\sqrt{2}\pi^{3\over 2}L^4}}
[(LM)^2 +\pi^2(LT)^2]^{3\over 4}
\e^{-\sqrt{(LM)^2+\pi^2(LT)^2}}
\cos[2\pi(Qa-{\delta\over 2})],
\end{equation}
from which one sees that there is no cubic terms. 
One sees that there is no cubic terms with respect to $a$ for 
fermion.

Let us turn to the cubic term again. 
By using the expansion formula for the boson field, 
the cubic terms coming from the bosonic field is given by
\begin{equation}
LV_{{\rm eff}}\bigg|_{cubic}=LV_{{\rm eff}}^{T\neq 0}\bigg|_{cubic}
=-\left({g_4\over{2\pi}}\right)^3 T E^{(3)}\phi^3,
\label{cubicboson}
\end{equation}
where we have defined 
$$
\phi\equiv {{2\pi}\over g_4}{a\over L}
$$
and the coefficient of the cubic term is given by 
\begin{equation}
E^{(3)}\equiv {\pi^2\over{48}}\left((3\times 40)
+(dN_{adj}^{(+)s}\times 40)
+(2N_{fd}^{(+)s}\times 8)
\right). 
\label{cubicboson2}
\end{equation}
The first term in Eq. (\ref{cubicboson2}) comes from the five dimensional
gauge field, and the second (third) does from the adjoint
(fundamental) scalar with the $\eta=+$ parity. We notice that
the cubic term entirely comes from the boson field with zero mode.
\par
If we employ the useful expression for the effective potential 
for the standard model \cite{baryon} in high temperature 
expansion, the strong first order phase transition is realized if 
\begin{equation}
{\phi(T_c)\over T_c} > 1,
\label{strong}
\end{equation}
where the VEV at the critical temperature is given by
\begin{equation}
\phi(T_c)\equiv v(T_c)={{2E^{(3)}T_c}\over \lambda}.  
\label{baryon}
\end{equation}
Here, the Higgs quartic coupling $\lambda$ depends
on temperature, but we assume its dependence is almost negligible. 
One obtains the upper bound of the Higgs mass
from the inequality (\ref{strong}). Since the 
cubic term for the
case of the minimal standard model is given by
$$
V_{SM}\bigg|_{cubic}=-E^{(3)}_{SM}T\phi^3,
$$
where
$E^{(3)}_{SM}=(2M_W^3 + M_Z^3)/4\pi v^3$. We obtain that 
\begin{equation}
\lambda < 2E^{(3)}_{SM},
\label{baryo}
\end{equation}
which implies that the Higgs mass is too small 
$m_H(=\sqrt{2\lambda}~v) \rough{<} 45$ GeV, 
which is inconsistent with the present lower 
bound on the Higgs mass.
\par
Let us apply the useful relation (\ref{strong}) to our case, 
where the coefficient $E^{(3)}$ is given by (\ref{cubicboson2}). 
We immediately understand the importance of the higher
dimensional gauge field to realize the strong first order 
phase transition. If we 
assume $N_{adj}^{(+)}=N_{fd}^{(+)}=0$, only the gauge 
field contributes to the cubic term. We 
obtain, from (\ref{cubicboson2}), that
$$
\lambda < g_4^3 \times {5\over {8\pi}},
$$
which implies that
$$
m_H~\rough{<}~155~~\mbox{GeV}\quad\mbox{for}\quad g_4\simeq {\cal O}(1). 
$$
It should be emphasized that the factor $3$ in 
Eq. (\ref{cubicboson2}), which stands for the 
on-shell degrees of freedom of the five dimensional gauge field,
is important to enhance the upper bound of the Higgs mass.
In fact, if we do not take into account the factor $3$, 
the upper bound on the Higgs mass is decreased as  
$$
m_H~\rough{<}~63~~\mbox{GeV}\quad\mbox{for}\quad g_4\simeq {\cal O}(1). 
$$
As in finite temperature phase transition in four dimensions,
the scalar field (with $\eta=+$), which has a zero mode, is
important for the strong first order phase transition. 
Therefore, if we introduce the scalar field with $\eta=+$ parity, 
we expect the phase transition to be stronger first order. 
This expectation is confirmed in our numerical studies.

We introduce the bulk mass terms in the model. 
In the high temperature expansion $LT \gg 1$, 
the cubic term is not directly affected by the bulk mass. 
But the bulk mass term yields the nonzero curvature 
at the origin of the effective potential, 
so that the bulk mass in general weakens the first 
order phase transition. The fermion field also tends 
to weaken the first order phase transition as well. 
\par
\begin{table}[t]
$$
\begin{array}{|c|cccc|cccccccc|}
\hline
 & z_{adj}^{(+)} & z_{fd}^{(+)} & z_{adj}^{(+)s} & z_{fd}^{(+)s} &  
   z_{adj}^{(-)} & z_{fd}^{(-)}   & z_{adj}^{(-)s} & z_{fd}^{(-)s}& 
   z^* & T_c & a_{T_c} & v(T_c)/T_c 
\\ \hline\hline
(1)& 0 & 0 & \mbox{-} & \mbox{-} & 0 & 0 & 0 & \mbox{-} & 0.018
& 114.5 & 0.0176&0.96 
\\ \hline
(2)& 0.1 & 0.2 & \mbox{-} & \mbox{-} & 0 & 0 & 0 & \mbox{-} & 0.011
& 134.0 & 0.0096&0.89
\\ \hline
(3)& 0 & 0 & \mbox{-} & \mbox{-} & 0 & 0 & 0.5 & \mbox{-} & 0.015
& 101.0 & 0.0139&0.90
\\ \hline
(4)& 0.1 & 0.2 & \mbox{-} & \mbox{-} & 0 & 0 & 0.2 & \mbox{-} & 0.010
& 137.1 & 0.0086&0.88
\\ \hline
(5)& 0.1 & 0.2 & \mbox{-} & \mbox{-} & 0.1 & 0 & 0 & \mbox{-} & 0.012
& 129.3 & 0.0114&0.91
\\ \hline
(6)& 0.1 & 0.2 & \mbox{-} & \mbox{-} & 0 & 0.2 & 0 & \mbox{-} & 0.012
& 130.7 & 0.0108&0.91
\\ \hline
(7)& 0.2 & 0.2 & \mbox{-} & \mbox{-} & 0.3 & 0.3 & 0 & \mbox{-} & 0.006
& 135.7 & 0.0068&1.19
\\ \hline
\end{array}
$$
\caption{Case (A) at finite temperature. 
$z^* \equiv R T_c$ where $T_c$ is the critical temperature. 
$a_{T_c}$ denotes the Higgs VEV at critical temperature.}
\label{AT}
$$
\begin{array}{|c|cccc|cccccccc|}
\hline
 & z_{adj}^{(+)} & z_{fd}^{(+)} & z_{adj}^{(+)s} & z_{fd}^{(+)s} &  
   z_{adj}^{(-)} & z_{fd}^{(-)}   & z_{adj}^{(-)s} & z_{fd}^{(-)s}& 
   z^* & T_c & a_{T_c}&v(T_c)/T_c 
\\ \hline\hline
(1)& 0 & 0 & \mbox{-} & 0 & 0 & 0 & 0 & \mbox{-} & 0.025
& 100.3 & 0.0309&1.22 
\\ \hline
(2)& 0.2 & 0.2 & \mbox{-} & 0.2 & 0 & 0 & 0 & \mbox{-} & 0.012
& 127.3 & 0.0135&1.10
\\ \hline
(3)& 0.2 & 0.1 & \mbox{-} & 0.2 & 0 & 0 & 0 & \mbox{-} & 0.015
& 118.7 & 0.0169&1.10
\\ \hline
(4)& 0.2 & 0.25 & \mbox{-} & 0.2 & 0 & 0 & 0 & \mbox{-} & 0.009
& 135.0 & 0.0099&1.13
\\ \hline
(5)& 0.2 & 0.3 & \mbox{-} & 0.5 & 0 & 0 & 0 & \mbox{-} & 0.016
& 122.2 & 0.0180&1.10
\\ \hline
\end{array}
$$
\caption{Case (B) at finite temperature}
\label{BT}
\end{table}
\begin{figure}[t]
\begin{center}
\includegraphics[width=9cm,height=9cm,keepaspectratio]
{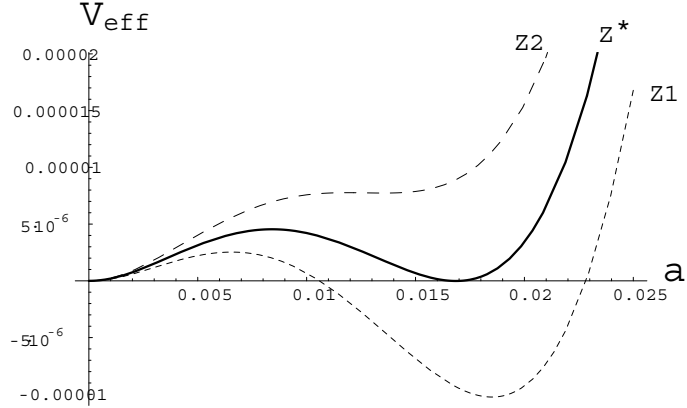}
\label{fig}
\end{center}
\caption{The behavior of the effective potential for the case (3)
in the Table $4$. The critical temperature is given by
$z^*=0.015345$ and $z1(z2)=0.01518(0.0155)$.}
\label{disegno1}
\end{figure}

We list the numerical result in the Table \ref{AT}.
The critical temperature (measured in GeV) is obtained by 
\begin{equation}
T_c\equiv {z^*\over R}=z^*\left({v\over a_0}\right)g_4 
\sim z^*\left({v\over a_0}\right)~~\mbox{for}~~g_4\sim O(1),
\end{equation}
where $v\simeq 246$ GeV and $a_0$ is determined as the minimum of
the zero temperature effective potential. The critical values of 
$a_{T_c}$ is obtained numerically and it yields 
\begin{equation}
{v(T_c)\over T_c}={1\over T_c}{a_{T_c}\over {g_4 R}}
={1\over g_4}{{a_{T_c}}\over z^*}\sim {{a_{T_c}}\over z^*}
~~\mbox{for}~~g_4\sim O(1).
\end{equation}  
We observe that the bulk mass actually weakens the first 
order phase transition. If we compare the 
cases $(2)$-$(6)$ with the case $(1)$, 
we see that the quantity $v(T_c)/T_c$ for $(2)$-$(6)$ 
is smaller than that of the case $(1)$. 
This shows that massive fields tend to weaken 
the first order phase transition, which is 
consistent with the common belief. 
\par
The result of the case $(7)$ is somewhat exceptional case. 
All the field but the gauge field is essentially massive in this
case, so that we do not expect the strong first order phase
transition. Due to the slightly large bulk mass, however, the 
fields with such large bulk mass do not contribute 
to the effective potential because of the Boltzmann 
like suppression factor in the modified Bessel 
function, $K_{5/2}({M\over{T}}\sqrt{(nLT)^2+l^2})$.
The five dimensional gauge fields dominates in the effective potential
and gives the main contribution to the cubic term. That's why
we have the strong first order phase transition. 
As we have explained in the previous paper \cite{MT2}, 
the Higgs mass in this case is lighter because of the $z^2$-suppression 
in the logarithmic factor.

As for the results in Table \ref{BT}, 
we can also see that the bulk mass weaken the first order phase transition 
by comparing the results (2)-(5) with that of (1). 
However, thanks to the scalar fields with periodic boundary conditions 
(specified by the flavor number $N_{fd}^{(+)s}$), the
coefficients of the cubic term of the Higgs potential is enhanced. 
Thus, we can still obtain the strong first order phase transition 
although the bulk mass weakens the strength of the first order
phase transition. This is the crucial difference 
between the case (A) and the case (B). 
In Figure \ref{disegno1}, we depict the behavior of the effective potential 
at finite temperature for the case (3) in the Table \ref{BT}.\footnote{Other 
cases also indicate similar behaviors.} 
And this is a typical behavior for the first order phase
transition obtained in the analyses.

We also find that there is an interesting 
behavior in this case. For the 
case (2), we have $m_H\simeq 118.58 (\sim 118.6)$ GeV 
with $v(T_c)/T_c\simeq
1.0949 (\sim 1.10)$. On the other hand, for the case (3), we have 
$m_H\simeq 119.63 (\sim 120)$ GeV with $v(T_c)/T_c\simeq
1.0985(\sim 1.10)$. Even though it is a tiny difference, contrary 
to the common sense, the heavier Higgs really induces the stronger 
first order phase transition. This is also considered to be the effect 
of the bulk mass on the phase transition.

\section{Conclusion}
In this paper, we have investigated the bulk mass effects on 
the electroweak phase transition at finite temperature 
in the five dimensional SU(3) gauge-Higgs unification theory, 
in particular the compatibility of the Higgs mass 
satisfying the experimental lower bound and the strong first order phase 
transition necessary for a successful electroweak baryogenesis. 

It is known that in general the massive matter and fermion 
tend to weaken the first order phase transition, so that
we expect the bulk mass does not enhance the first order phase
transition. This is actually the case in our numerical studies.
If the model contains the zero mode for scalar field 
like in the case (B), we can still have the strong 
first order phase transition. 


Our numerical results in this paper show that the matter content of 
the case (B) is favorable, where the Higgs mass 
is above the lower bound and the electroweak baryogenesis works well. 
This is the reason why the scalar fields with periodic boundary condition 
included only in the case (B) give the contributions for the cubic term 
in the effective potential and enhance the strength of 
the first order phase transition. 
From the experimental point of view, the case (B) is also interesting 
in that the Higgs mass is predicted to be close to the lower bound. 

In the standard argument of the Higgs mass and the 
order of the phase transition, it is known that 
the lighter Higgs tends to give the stronger first 
order phase transition. This is also understood from
Eqs. (\ref{strong}) and (\ref{baryon}).
In our examples studied in this paper, namely, the 
cases (2) and (3) in (B), we found an interesting behavior 
that the heavier Higgs gives the stronger first order phase
transition. This behavior helps us to construct realistic models.


In this paper, we have not addressed another important issue 
in gauge-Higgs unification, namely the problem of too small top mass. 
In gauge-Higgs unification, Yukawa coupling is dictated 
by the gauge coupling. 
Therefore, it is nontrivial to generate the order one coupling, 
such as the top Yukawa coupling. 
Recent proposal on this issue were made by \cite{CCP} 
in terms of large representation fermions and \cite{PSW} 
in terms of an explicit violation of 5D Lorenz invariance. 
It would be very interesting to examine whether the top mass 
can be obtained 
without spoiling the good points obtained in our analysis. 
This is left for a future work. 

\section*{Acknowledgements}
N.M. is supported by INFN, Sezione di Roma. 
K.T. is supported by the 21st Century COE Program at Tohoku
University.



\begin{thebibliography}{100}
%
%
\bibitem{gaugehiggs1}
N.~S.~Manton,
  Nucl.\ Phys.\ B {\bf 158}, 141 (1979);
D.~B.~Fairlie,
  Phys.\ Lett.\ B {\bf 82}, 97 (1979).
%
\bibitem{gaugehiggs2}
  N.~V.~Krasnikov,
  Phys.\ Lett.\ B {\bf 273} (1991) 246;
N.~Arkani-Hamed, A.~G.~Cohen and H.~Georgi,
  Phys.\ Lett.\ B {\bf 513}, 232 (2001); 
G.~R.~Dvali, S.~Randjbar-Daemi and R.~Tabbash,
  Phys.\ Rev.\ D {\bf 65}, 064021 (2002); 

\bibitem{HIL}
H.~Hatanaka, T.~Inami and C.~S.~Lim,
  Mod.\ Phys.\ Lett.\ A {\bf 13}, 2601 (1998);

\bibitem{ABQ}
I.~Antoniadis, K.~Benakli and M.~Quiros,
  New J.\ Phys.\  {\bf 3}, 20 (2001). 
  
%
%
\bibitem{hosotani}
Y.~Hosotani,
  Phys.\ Lett.\ B {\bf 126}, 309 (1983), 
  Annals Phys.\  {\bf 190}, 233 (1989).
%
%
\bibitem{gaugehiggs3}
  K.~Takenaga,
  Phys.\ Rev.\ D {\bf 64}, 066001 (2001);
  Phys.\ Rev.\ D {\bf 66}, 085009 (2002); 
  L.~J.~Hall, Y.~Nomura and D.~R.~Smith,
  Nucl.\ Phys.\ B {\bf 639}, 307 (2002); 
  M.~Kubo, C.~S.~Lim and H.~Yamashita,
  Mod.\ Phys.\ Lett.\ A {\bf 17}, 2249 (2002); 
 C.~Csaki, C.~Grojean and H.~Murayama,
  Phys.\ Rev.\ D {\bf 67}, 085012 (2003); 
  G.~Burdman and Y.~Nomura,
  Nucl.\ Phys.\ B {\bf 656}, 3 (2003); 
 N.~Haba and Y.~Shimizu,
  Phys.\ Rev.\ D {\bf 67}, 095001 (2003)
  [Erratum-ibid.\ D {\bf 69}, 059902 (2004)]; 
C.~A.~Scrucca, M.~Serone and L.~Silvestrini,
  Nucl.\ Phys.\ B {\bf 669}, 128 (2003); 
  K.~W.~Choi, N.~Haba, K.~S.~Jeong, K.~i.~Okumura, 
  Y.~Shimizu and M.~Yamaguchi,
  JHEP {\bf 0402}, 037 (2004); 
 N.~Haba, Y.~Hosotani, Y.~Kawamura and T.~Yamashita,
  Phys.\ Rev.\ D {\bf 70}, 015010 (2004); 
  N.~Haba, K.~Takenaga and T.~Yamashita,
  Phys.\ Rev.\ D {\bf 71}, 025006 (2005); 
G.~Martinelli, M.~Salvatori, C.~A.~Scrucca and L.~Silvestrini,
  JHEP {\bf 0510}, 037 (2005); 
  I.~Gogoladze, T.~Li, Y.~Mimura and S.~Nandi,
  Phys.\ Rev.\ D {\bf 72}, 055006 (2005); 
 N.~Haba, S.~Matsumoto, N.~Okada and T.~Yamashita,
  JHEP {\bf 0602}, 073 (2006).

\bibitem{CCP}
G.~Cacciapaglia, C.~Csaki and S.~C.~Park,
  JHEP {\bf 0603}, 099 (2006); 

\bibitem{PSW}
G.~Panico, M.~Serone and A.~Wulzer,
  Nucl.\ Phys.\ B {\bf 739}, 186 (2006); 



\bibitem{warped}
R.~Contino, Y.~Nomura and A.~Pomarol,
  Nucl.\ Phys.\ B {\bf 671}, 148 (2003); 
K.~Agashe, R.~Contino and A.~Pomarol,
  Nucl.\ Phys.\ B {\bf 719}, 165 (2005); 
  K.~y.~Oda and A.~Weiler,
  Phys.\ Lett.\ B {\bf 606}, 408 (2005); 
 Y.~Hosotani and M.~Mabe,
  Phys.\ Lett.\ B {\bf 615}, 257 (2005); 
Y.~Hosotani, S.~Noda, Y.~Sakamura and S.~Shimasaki,
  Phys.\ Rev.\ D {\bf 73}, 096006 (2006). 
%
\bibitem{HTY}
K.~Takenaga,
  Phys.\ Lett.\ B {\bf 570}, 244 (2003); 
  N.~Haba, K.~Takenaga and T.~Yamashita,
  Phys.\ Lett.\ B {\bf 605}, 355 (2005). 



\bibitem{PS}
G.~Panico and M.~Serone,
  JHEP {\bf 0505}, 024 (2005). 

\bibitem{MT}
N.~Maru and K.~Takenaga,
  Phys.\ Rev.\ D {\bf 72}, 046003 (2005). 

\bibitem{MT2}
N.~Maru and K.~Takenaga,
  Phys.\ Lett.\ B {\bf 637}, 287 (2006). 



\bibitem{MY}
  N. Maru and T. Yamashita, arXiv:hep-ph/0603237. 

\bibitem{LMH}
  C.~S.~Lim, N.~Maru and K.~Hasegawa,
  arXiv:hep-th/0605180.

\bibitem{GIQ}
G.~von Gersdorff, N.~Irges and M.~Quiros,
Nucl.\ Phys.\ B {\bf 635}, 127 (2002). 

\bibitem{hosont}
Y.~Hosotani, S.~Noda and K.~Takenaga,
Phys.\ Lett.\ B {\bf 607}, 276 (2005); 

%
\bibitem{HLM} 
  K.~Hasegawa, C.~S.~Lim and N.~Maru,
  Phys.\ Lett.\ B {\bf 604}, 133 (2004). 
%
\bibitem{yukawa}
 C.~Csaki, C.~Grojean and H.~Murayama;
  G.~Burdman and Y.~Nomura; 
C.~A.~Scrucca, M.~Serone and L.~Silvestrini in \cite{gaugehiggs3}. 




\bibitem{CW}
  S.~R.~Coleman and E.~Weinberg,
  Phys.\ Rev.\ D {\bf 7}, 1888 (1973).
%
%
\bibitem{FKT}
  K.~Funakubo, A.~Kakuto and K.~Takenaga,
  Prog.\ Theor.\ Phys.\  {\bf 91}, 341 (1994). 

\bibitem{DJ}
  L.~Dolan and R.~Jackiw,
  Phys.\ Rev.\ D {\bf 9}, 2904 (1974).


\bibitem{baryon}
A.~G.~Cohen, D.~B.~Kaplan and A.~E.~Nelson,
  Ann.\ Rev.\ Nucl.\ Part.\ Sci.\  {\bf 43}, 27 (1993); 
K.~Funakubo,
  Prog.\ Theor.\ Phys.\  {\bf 96}, 475 (1996); 
V.~A.~Rubakov and M.~E.~Shaposhnikov,
  Usp.\ Fiz.\ Nauk {\bf 166}, 493 (1996)
  [Phys.\ Usp.\  {\bf 39}, 461 (1996)]. 

\end{thebibliography}
\end{document}